\documentclass[article,aps, twocolumn,showpacs,amsmath,amssymb,superscriptaddress,nofootinbib,floatfix,prl]{revtex4-1}

\usepackage[utf8]{inputenc}
\usepackage{color}         
\usepackage[pdftex]{graphicx}      
\usepackage{bm}
\usepackage{natbib}            
\usepackage[colorlinks,plainpages=false,linkcolor=black,urlcolor=black,citecolor=black,pdfpagemode=UseNone,pdfstartview=FitBH]{hyperref}
\usepackage{float}

\begin{document}

\title{Self-stacked 1$\mathrm{T}$-1$\mathrm{H}$ layers in 6$\mathrm{R}$-NbSeTe and the emergence of charge and magnetic correlations due to ligand disorder}

\author{S. K. Mahatha}
\thanks{These authors contributed equally to this work.}

\affiliation{UGC-DAE Consortium for Scientific Research, University Campus, Khandwa Road, Indore-452001, India}
\affiliation{Ruprecht Haensel Laboratory, Deutsches Elektronen-Synchrotron DESY, 22607 Hamburg, Germany}

\author{J. Phillips}
\thanks{These authors contributed equally to this work.}
\affiliation{Departamento de Física Aplicada, Universidade de Santiago de Compostela, E-15782 Campus Sur s/n, Santiago de Compostela, Spain}
\affiliation{Instituto de Materiais iMATUS, Universidade de Santiago de Compostela, E-15782 Campus Sur s/n, Santiago de Compostela, Spain} 

\author{J. Corral-Sertal}
\thanks{These authors contributed equally to this work.}
\affiliation{Departamento de Física Aplicada, Universidade de Santiago de Compostela, E-15782 Campus Sur s/n, Santiago de Compostela, Spain}
\affiliation{CiQUS, Centro Singular de Investigacion en Quimica Biolóxica e Materiais Moleculares, Departamento de Quimica-Fisica, Universidade de Santiago de Compostela, Santiago de Compostela, E-15782, Spain.}

\author{D. Subires}
\affiliation{Donostia International Physics Center (DIPC), San Sebastián, Spain}

\author{A. Korshsunov}
\affiliation{European Synchrotron Radiation Facility (ESRF), BP 220, F-38043 Grenoble Cedex, France}

\author{A. Kar}
\affiliation{Donostia International Physics Center (DIPC), San Sebastián, Spain}

\author{J. Buck}
\affiliation{Ruprecht Haensel Laboratory, Deutsches Elektronen-Synchrotron DESY, 22607 Hamburg, Germany}
\affiliation{Institut f\"{u}r Experimentelle und Angewandte Physik, Christian-Albrechts-Universit\"{a}t zu Kiel, D-24098 Kiel, Germany}

\author{F. Diekmann}
\affiliation{Ruprecht Haensel Laboratory, Deutsches Elektronen-Synchrotron DESY, 22607 Hamburg, Germany}
\affiliation{Institut f\"{u}r Experimentelle und Angewandte Physik, Christian-Albrechts-Universit\"{a}t zu Kiel, D-24098 Kiel, Germany}

\author{Y. P. Ivanov}
\affiliation{Electron Spectroscopy and Nanoscopy, Istituto Italiano di Tecnologia, Via Morego 30, 16163 Genova, Italy}

\author{A. Chuvilin}
\affiliation{CIC Nanogune, San Sebastián, Spain}
\affiliation{IKERBASQUE, Basque Foundation for Science, 48013 Bilbao, Spain}

\author{D. Mondal}
\affiliation{Consiglio Nazionale delle Ricerche (CNR)- Istituto Officina dei Materiali (IOM), Laboratorio TASC in Area Science, Trieste, Italy}
\affiliation{ Sovarani Memorial College, Jagatballavpur, Howrah - 711408, India }                

\author{I. Vobornik}
\affiliation{Consiglio Nazionale delle Ricerche (CNR)- Istituto Officina dei Materiali (IOM), Laboratorio TASC in Area Science, Trieste, Italy}

\author{A. Bosak}
\affiliation{European Synchrotron Radiation Facility (ESRF), BP 220, F-38043 Grenoble Cedex, France}

\author{K. Rossnagel}
\affiliation{Ruprecht Haensel Laboratory, Deutsches Elektronen-Synchrotron DESY, 22607 Hamburg, Germany}
\affiliation{Institut f\"{u}r Experimentelle und Angewandte Physik, Christian-Albrechts-Universit\"{a}t zu Kiel, 24098 Kiel, Germany}

\author{V. Pardo}
\affiliation{Departamento de Física Aplicada, Universidade de Santiago de Compostela, E-15782 Campus Sur s/n, Santiago de Compostela, Spain}
\affiliation{Instituto de Materiais iMATUS, Universidade de Santiago de Compostela, E-15782 Campus Sur s/n, Santiago de Compostela, Spain} 

\author{Adolfo O. Fumega}
\affiliation{Department of Applied Physics, Aalto University, 02150 Espoo, Finland}

\author{S. Blanco-Canosa}
\email{sblanco@dipc.org}
\affiliation{Donostia International Physics Center (DIPC), San Sebastián, Spain}
\affiliation{IKERBASQUE, Basque Foundation for Science, 48013 Bilbao, Spain}
\date{February 2024}

\begin{abstract}
The emergence of correlated phenomena arising from the combination of 1$\mathrm{T}$ and 1$\mathrm{H}$ van der Waals layers is the focus of intense research. 
Here, we synthesize a novel self-stacked 6$\mathrm{R}$ phase in NbSeTe, showing a perfect alternating 1T and 1H layers that grow coherently along the c-direction, as revealed by scanning transmission electron microscopy. Angle resolved photoemission spectroscopy shows a mixed contribution of the trigonal and octahedral Nb bands to the Fermi level. Diffuse scattering reveals temperature-independent short-range charge fluctuations with propagation vector $\mathrm{q_{CO}}$=(0.25,0), derived from the condensation of a longitudinal mode in the 1T layer. We observe that ligand disorder quenches the formation of a charge density wave.
Magnetization measurements suggest the presence of an inhomogeneous, short-range magnetic order, further supported by the absence of a clear phase transition in the specific heat. These experimental analyses in combination with \textit{ab initio} calculations indicate that the ground state of 6$\mathrm{R}$-NbSeTe is described by a statistical distribution of short-range charge-modulated and spin-correlated regions driven by ligand disorder. 
Our results devise a route to synthesize 1$\mathrm{T}$-1$\mathrm{H}$ self-stacked bulk heterostructures to study emergent phases of matter.
\end{abstract}

\maketitle
 
\section{Introduction}
Layered van der Waals (vdW) materials consisting of the stacking of two-dimensional (2D) transition metal dichalcogenide (TMD) layers offer a fertile playground to realize novel physical phenomena \cite{vdwHT2013,doi:10.1126/science.aac9439,Zeng2018}. The great versatility of the TMD's chemical structure and the presence of a vdW gap allows the intercalation of a large variety of atomic species, offering endless possibilities for materials engineering \cite{manzeli20172d}, emergence of correlated phases or tuning the balance between different competing orders that intertwine with the ground state of the vdW layers \cite{Nair2020,Takagi2023,Wu2023,Yu2015,PhysRevLett.109.176403,https://doi.org/10.1002/adma.201801325,doi:10.1126/sciadv.abb9379,PhysRevMaterials.7.064401}. 

The most common strategy to engineer vdW heterostructures is a bottom-up approach where single layers are exfoliated and stacked together forming few-layer slabs \cite{https://doi.org/10.1002/adfm.202007810,li2017epitaxial}.
This approach takes advantage of the unique degrees of freedom encountered in vdW materials, such as the twist angle between layers or the combination of layers with different symmetry-breaking orders to design and tune strongly correlated electronic phases \cite{Cao2018,Cao2018a,shabani2021deep,Zeng2023,Cai2023,Kezilebieke2020,Vano2021}.

Nevertheless, in bulk TMDs, the growth and stacking of different layers turns out to be experimentally demanding, and limited to the metastable  4$\mathrm{H}_b$ phase of TaS$_2$ \cite{di1973preparation}. 
This vdW heterostructure benefits from the stacking of the strongly correlated insulating 1$\mathrm{T}$ layers, featuring nearly flat bands \cite{wang2020band}, and the metallic 1$\mathrm{H}$ layer to promote chiral superconductivity and topological edge modes \cite{Ribak2020,Nayak2021}. A similar 6$\mathrm{R}$-phase, stacking 1$\mathrm{T}$-1$\mathrm{H}$ layers of TaS$_2$ hosting a charge density wave (CDW), becomes superconducting with higher T$_c$ than the parent compound \cite{achari2022alternating}.
Therefore, achieving an alternating 1$\mathrm{T}$-1$\mathrm{H}$ bulk heterostructure in systems other than TaS$_2$ would expand the possibilities to explore the complex phase diagram of correlated phases in TMDs.

\begin{figure*}
\begin{center}
\includegraphics[width=\textwidth,draft=false]{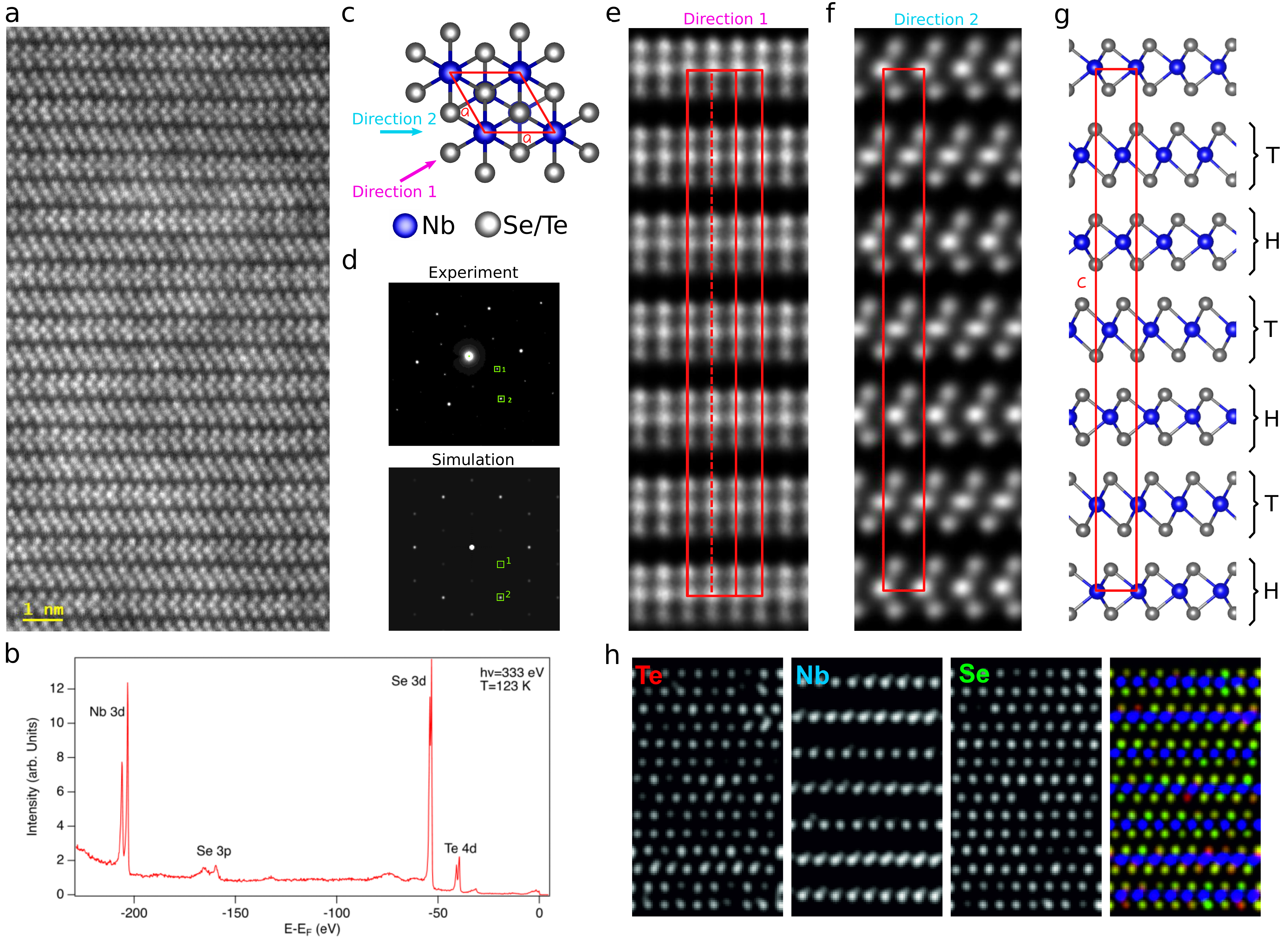}
\caption{(a) Large area ADF-STEM image of the sample showing the coherent growth of a layered van der Waals structure. (b) Core level x-ray photoemission, showing the principal emission lines of Nb, Se, and Te.
(c) Top view of the 6$\mathrm{R}$-NbSeTe unit cell showing the directions of the ADF-STEM lateral images (e) and (f). (d) Experimental electron diffraction patterns in 6$\mathrm{R}$-NbSeTe in the 001 direction and simulation considering a stochastic distribution of the ligands atoms. 
(e-f) ADF-STEM lateral images highlighting the 1$\mathrm{T}$-1$\mathrm{H}$ heterobilayer in the 6$\mathrm{R}$-NbSeTe. A 1/3,-1/3) in-plane displacement of the 1$\mathrm{T}$-1$\mathrm{H}$ bilayer can be seen in (f) and in the lateral view of the conventional unit cell in (g). 
(h) EDX atomic resolution maps of the image shown in (f) reveal that Se and Te are distributed stochastically (images processed for noise reduction). 
}
\label{Fig1}
\end{center}
\end{figure*}

Here, we present an alternative route to obtain 1$\mathrm{T}$-1$\mathrm{H}$ bulk structures, and in particular the 6$\mathrm{R}$-phase of NbSeTe. This kind of heterostructure can be synthesized by the chemical substitution of the chalcogen atom (S, Se, or Te) from the parent 1$\mathrm{T}$-NbTe$_2$ and 2$\mathrm{H}$-NbSe$_2$ bulk phases. The chemical ligand-doping of the parent compounds opens the possibility of tuning lattice symmetries and artificially creating new electronic ground states. For instance, the distorted quasi-1D 1$\mathrm{T}$ structure of NbTe$_2$ is characterized by a first-order-like CDW transition \cite{Battaglia2005}, while the 2$\mathrm{H}$ polytype of NbSe$_2$ develops at low temperature a superconducting ground state coexisting with CDW order \cite{lian2018unveiling,Du2000,Borisenko2009}. Given the extreme versatility of the vdW TMD fabric and the proximity of collective phenomena with similar energy scales, the random substitution of the TMD ligand offers new avenues to engineer novel ground states and the possibility of targeted functionalities. For instance, recently, the metastable layered compound 1$\mathrm{T}$-NbSeTe has been synthesized \cite{yan2019nbsete}, highlighting the first pure 1$\mathrm{T}$ superconducting phase within the TMD family. Furthermore, the chemical substitution of Te by Se in the solid solution TaSe$_{2-x}$Te$_x$ (already superconducting in bulk form \cite{luo2015polytypism,liu2016nature}) has been predicted to give rise to emergent phases and topological superconductivity at the monolayer limit \cite{phillips2023self}. 

\begin{figure}[ht!]
\begin{center}
\includegraphics[width=0.99\columnwidth,draft=false]{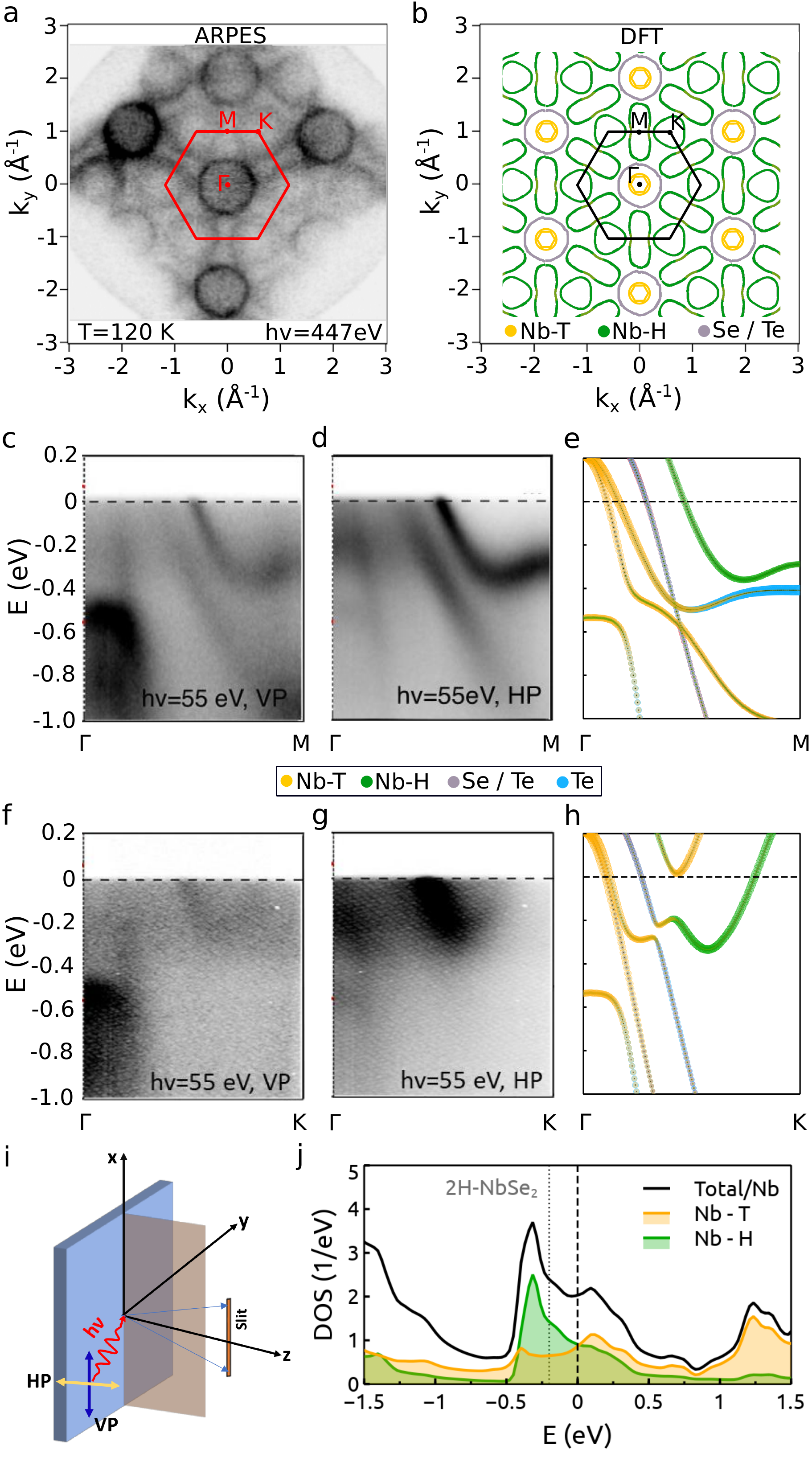}
	\caption{(a) Fermi surface map of the NbSeTe single crystal at $h\nu$= 447 eV, T= 120 K. (b) Calculated Fermi surfaces using DFT. The different colors indicate the dominant contribution of each sheet in the Fermi surface. (c)-(d) $\Gamma$-M normal emission ARPES spectra of NbSeTe and the calculated band structure obtained from DFT (e) for a similar energy interval around the Fermi level and also in the $\Gamma$-M direction. (f)-(g) $\Gamma$-K Normal emission ARPES spectra of NbSeTe along with DFT calculated band structure (h). The ARPES spectra were obtained with 55 eV incident photon energy with vertical (c)-(f) and horizontal (d)-(g) polarizations. (i)ARPES experimental geometry with the linear horizontal polarized (HP) and vertical polarized (VP) light vectors. (j) Calculated density of states showing the contribution coming from the different Nb atoms in the structure, namely those in T- and H- layers. The dotted vertical line at -0.20 eV shows the location of the Fermi level for 2H-NbSe$_2$ in the projected DOS of Nb-H in 6$\mathrm{R}$-NbSeTe.}
	\label{Fig2}
 \end{center}
\end{figure}

In this work, we tweak the growth parameters to chemically combine 2$\mathrm{H}$-NbSe$_2$ and 1$\mathrm{T}$-NbTe$_2$ producing a 6$\mathrm{R}$-phase of bulk NbSeTe, described by a perfect 3$\times$stacking of c-axis oriented 1$\mathrm{T}-$1$\mathrm{H}$ heterobilayers. This rhombohedral 6$\mathrm{R}$ structure shows a decoupling of the vdW layers that weakens the charge correlations of the parent compounds and allows the emergence of a glassy-type short-range competing magnetic order. 
Remarkably, the short-range charge correlations appear as diffuse scattering with in-plane propagation vector $\mathrm{q_{CO}}$=(0.25,0) and are driven by the condensation of a phonon of the 1$\mathrm{T}$-Nb layer that strongly competes with a different charge modulation with propagation vector $\left(\mathrm{q^*}=\frac{1}{6},\,\frac{1}{6}\right)$ in the same 1$\mathrm{T}$-Nb slab. \textit{Ab initio} calculations show that this kind of charge ordering emerges independently of the in-plane Se/Te disorder pattern and presumably introduces local strains that drive local magnetic order. Our work describes a path to study hidden phases and demonstrates that the large versatility of the layered TMD structure allows tuning the structural, magnetic, and electronic properties to reveal novel and hidden phases of matter.

\section{Results}
Let us start by analyzing in detail the 6$\mathrm{R}$ structure of NbSeTe.
Figure \ref{Fig1}(a) demonstrates the layered growth of NbSeTe. Further, atomic resolution annular dark-field in the scanning transmission electron microscope (ADF-STEM) imaging, Fig. \ref{Fig1}(e) and (f), shows a perfect stacking of 1$\mathrm{T}$ and 1$\mathrm{H}$ layers, where we can identify and assign the Nb atoms sandwiched between the chalcogen layers that feature both a trigonal, 1$\mathrm{H}$, and octahedral, 1$\mathrm{T}$, coordination. The conventional unit cell of the 6$\mathrm{R}$-NbSeTe structure is shown in Figs. \ref{Fig1}(c) and \ref{Fig1}(g). 
The high-resolution transmission electron microscopy (TEM) also resolves an in-plane shift of consecutive bilayers of ($\frac{1}{3}$, -$\frac{1}{3}$) ( Fig. \ref{Fig1}(f)) in the unit cell defined in Fig. \ref{Fig1} (c), being the overall structure described by a rhombohedral primitive unit cell with space-group symmetry \textit{R3m} (no. 160). 
The alternating 3$\times$ 1T-1H consecutive heterobilayers, Fig. \ref{Fig1}(g), is confirmed by x-ray diffraction that provides the lattice parameters \textit{a}=\textit{b}= 3.53 \AA\ and $\mathrm{c}$= 39.26 \AA, for a Se/Te ratio of 0.91, as obtained from energy dispersive analysis (EDX).
Results of core level x-ray photoelectron spectroscopy (XPS) measurements are shown in Fig. \ref{Fig1}(b), presenting data obtained with an incident photon energy E$_\mathrm{in}$= 333 eV. It shows the emissions from the spin-orbit split bands of 3\textit{d} (4\textit{d}) orbitals of Nb and Se (Te) at 220 and 55 (40) eV below the Fermi level, respectively. Besides, the 3\textit{p} emissions are also observed at 160-166 eV, in good agreement with the XPS spectra of NbSe$_2$ and NbTe$_2$ and reports in the literature \cite{Wang2017}.
Figure \ref{Fig1}(f) shows the EDX atomic resolution maps after noise reduction of the structure, which show that Se and Te are stochastically distributed, discarding a Janus-type coherent growth. The disorder between ligands will have a major effect on the emergence of short-range charge and magnetic correlations.

Having structurally characterized the 6$\mathrm{R}$ phase of NbSeTe single crystals, we proceed with the study of its electronic structure. Figure \ref{Fig2}(a) displays the  experimental hexagonal Fermi surface obtained at 447 eV photon energy (see ARPES geometry in Fig. \ref{Fig2} (i)), which can be compared with the Fermi surface calculated using density functional theory (DFT) shown in Fig. \ref{Fig2}(b). It consists of two hole-like circular pockets around $\Gamma$ (mainly with anionic character) and $\mathrm{K}$ (of Nb($\mathrm{H}$) character) and one electron-like ``dogbone" centered at $\mathrm{M}$ (coming mainly from Nb(H) bands), resembling the 2$\mathrm{H}$-polymorph of TaSe$_2$ \cite{Rossnagel2011}. 
The additional hole pocket centered at $\Gamma$, which is absent in 2$\mathrm{H}$-TaSe$_2$, comes from Nb($\mathrm{T}$) layers (see high-resolution data in the Supplementary Information).    

We can further analyze the band structure in more detail in an energy region below the Fermi level. The ARPES results are presented in Fig. \ref{Fig2}(c-h) along the high symmetry directions together with the corresponding bands (with the main band character denoted by colors) obtained from DFT calculations. Qualitatively, the energy versus wavevector intensity maps match extremely well with the simulated bands, both along the $\Gamma$-M and $\Gamma$-K directions. 
Our results show no evidence of gap opening at the Fermi level, and Fermi surface nesting appears to be weak at $\mathrm{q_{CDW}}$=2/3$\overline{\Gamma\mathrm{M}}$ and $\mathrm{q_{CDW}}$=2/3$\overline{\Gamma\mathrm{K}}$. In Fig. \ref{Fig2}(j) we show the calculated density of states in a layer-resolved fashion, showing separately the contribution from Nb d-bands originating from T and H layers. We see that the anion p weight close to the Fermi level is relatively small, although some pieces of the Fermi level do come from anion p orbitals, with a larger Te character close to the Fermi level (the Te p bands will be typically higher in energy than the Se p bands) \cite{phillips2023self}.

\begin{figure}
\begin{center}
\includegraphics[width=\columnwidth,draft=false]{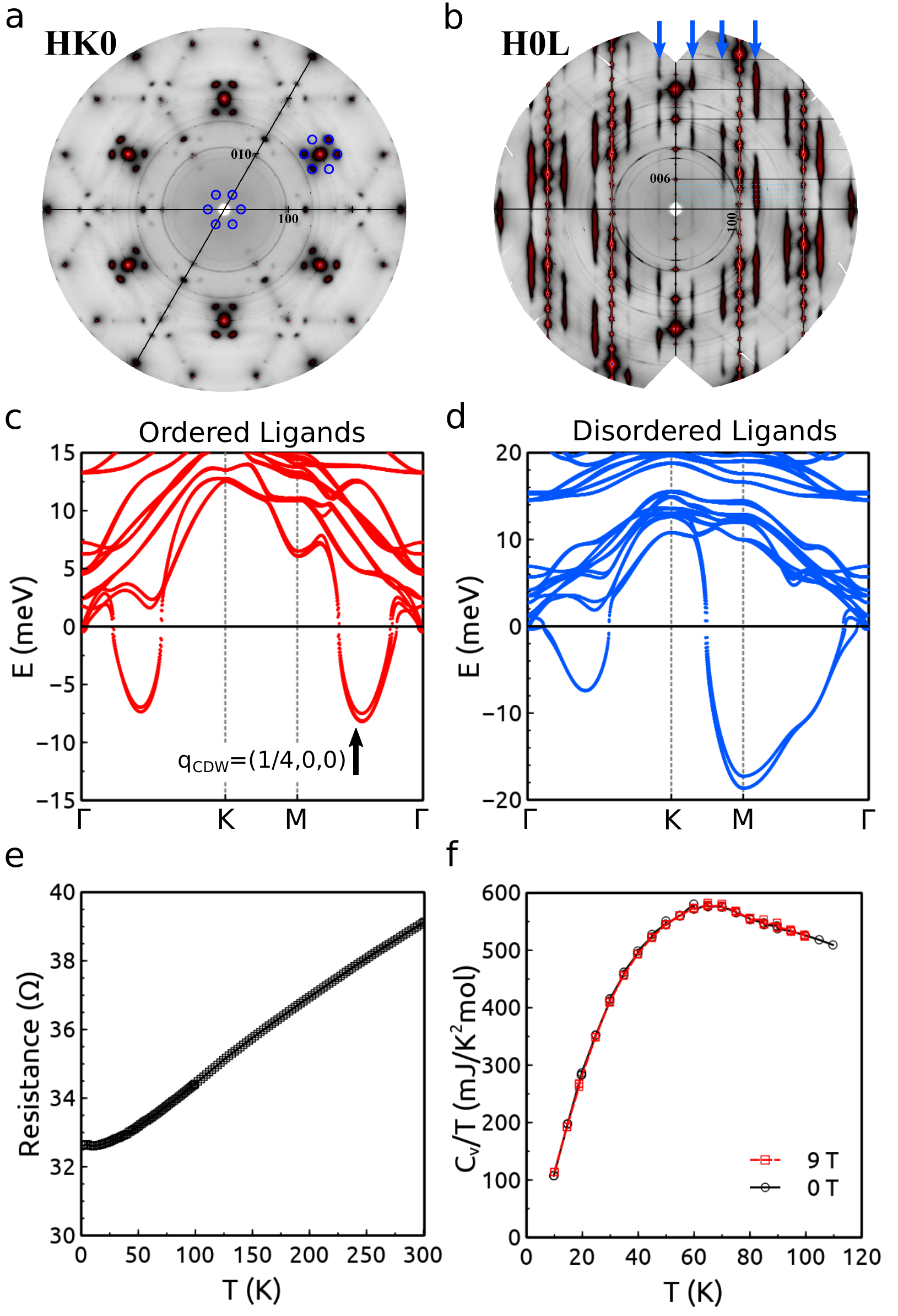}
\caption{(a) HK0 and (b) H0L diffuse scattering maps of NbSeTe at 80 K. Strong DS is observed in the vicinity of the 110 Bragg peak in (a) and as rods along the L-direction in (b), highlighted with blue circles and arrows respectively. (c-d) Harmonic phonon spectra for the structure having NbSe$_2$ (NbTe$_2$) with local 1H (1T) coordination (ordered ligands) and one with different ligands both in 1H and 1T coordinations (called disordered ligands). Two imaginary modes appear to be competing, but triply degenerate. Only the $\Gamma$ to $M$ mode is seen in our experiments, indicating the other one is hidden by the competition between possible orders. The disordered ligands plot shows the imaginary modes displaced towards the $\mathrm{M}$ point. (e) Temperature dependence of the resistivity for the NbTeSe single crystal. (f) Temperature-dependent heat capacity measurement at H =0 T and 9 T.
}
\label{Fig3}
\end{center}
\end{figure}

We note that the absence of clear translation vectors that may nest portions of the Fermi surface does not preclude the presence of charge modulations. Indeed, this is a common occurrence in TMDs, dominated by electron-phonon interactions for the CDW formation \cite{Diego2021,Weber2018,doi:10.1021/acs.nanolett.2c04584,PhysRevB.109.035133}.
This is illustrated in Fig. \ref{Fig3}, where we present the diffuse scattering (DS) data. First, we observe strong diffuse rods of intensity in between Bragg peaks along the L direction, characteristic of a coherent growth of a 2D layered structure, see Fig. \ref{Fig3}(b). Second, the diffuse scattering (DS) maps reveal the presence of a diffuse signal, characteristic of a precursor of charge modulations, in the $\mathrm{H K 0}$ plane, with in-plane propagation vector $\mathrm{q_{CO}}$=(0.25 0) r.l.u. We also notice that the DS develops a sizable intensity in the vicinity of the (1 1 0) Bragg reflection. This DS does not match either the propagation vector of the charge modulations of the bulk 2$\mathrm{H}$-NbSe$_2$ \cite{Weber2011} or 1$\mathrm{T}$-NbTe$_2$ \cite{Jang2022}. The diffuse clouds in the $\mathrm{H K 0}$ plane are indicative of a 0-dimensional structure without a fully ordered in-plane direction. The short-range correlation length of the charge fluctuations amounts to 3-4 nm in the \textit{a}-\textit{b} plane and does not vary with lowering the temperature down to 80 K, suggesting that the charge correlations are static and do not involve a soft mode. In the $\mathrm{H 0 L}$ plane, the DS is present as diffuse streaks of intensity along the \textit{c}-axis, Fig. \ref{Fig3}(b). The elongated rod in the DS indicates that the signal comes from in-plane atomic displacements perpendicular to the \textit{c}-direction, hence concentrating the charge fluctuations within a single Nb-TMD layer of the 3$\times$1$\mathrm{T}$-1$\mathrm{H}$ heterobilayer.   

In order to elucidate the presence of charge correlations in 6$\mathrm{R}$-NbSeTe, we have calculated the phonon dispersion of the ordered 1T(Te)-1H(Se) in the 6$\mathrm{R}$-NbSeTe structure [Fig. \ref{Fig3}(c)]. It shows three imaginary soft modes at the propagation vector we observe experimentally; i.e., $\mathrm{q_{CO}}$= $\frac{1}{2}\overline{\Gamma\mathrm{M}}$=(0.25, 0) r.l.u. The soft acoustic branches do not share any resemblance with the charge modulations in the parent phases of NbSe$_2$ and NbTe$_2$, which adopt 3$\times$3 CDW ordering at 33 K and 170 K, respectively. Besides, the scenario drawn to explain the CDW in the parent compounds, i.e. Fermi surface nesting, seems to fail in NbSeTe, where the short-range correlations and the absence of temperature dependence of the DS point to the order-disorder dynamics of the CDW. 
These triply degenerate soft modes correspond to each of the three 1$\mathrm{T}$-NbTe$_2$ layers of the 6$\mathrm{R}$-NbSeTe heterostructure and describe an in-plane vibration of the Nb atoms. Furthermore, the corresponding Nb atoms of the consecutive 1$\mathrm{H}$-NbSe$_2$ layer appear static; i.e., do not follow the 1$\mathrm{T}$-Nb displacement, hence the 1$\mathrm{T}$-Nb vibration is completely sandwiched between 1$\mathrm{H}$-Nb layers. The lattice dynamics are fully consistent with the observations of the strong DS perpendicular to the $\mathrm{H K}$ plane of Fig. \ref{Fig3}(b).
On the other hand, the phonon calculations also unveil a second instability at $\mathrm{q}$=($\frac{1}{6},\,\frac{1}{6}$) in the ${\Gamma-\mathrm{K}}$ path describing also the in-plane Nb vibration of the 1$\mathrm{T}$-NbTe$_2$, not observed in the DS experiments. The observation of only one of the instabilities hints at the competition between the two types of charge periodicities, akin to the high-T$_c$ cuprates \cite{Gerber2015,Kim2018}, kagome materials \cite{Korshunov2023,Cao2023,Binghai2023} or similar TMD van der Waals compounds\cite{doi:10.1021/acs.nanolett.2c04584}.

To mimic the effect of chemical disorder, we have exchanged the occupations and considered the Se (Te) atoms in the octahedral (trigonal) symmetry and vice versa. We found that, at the DFT level, the relaxed 6$\mathrm{R}$ structure consisting of alternating 1$\mathrm{T}$-NbTe$_2$-1$\mathrm{H}$-NbSe$_2$ layers is 0.1 eV/Nb atom more stable than the 1$\mathrm{H}$-NbTe$_2$-1$\mathrm{T}$-NbSe$_2$ structure. The solution with disordered ligands (of higher energy) shows a different type of imaginary modes, peaking in this case at the $\mathrm{M}$ point [see Fig. \ref{Fig3}(d)]. Therefore, this result suggests that ligand disorder decreases the coherence of the $\frac{1}{2}\overline{\Gamma\mathrm{M}}$-peaks, thus quenching the formation of a long-range CDW and establishing the short-range charge correlations observed experimentally.

Finally, Figs. \ref{Fig3}(e) and \ref{Fig3}(f) show the temperature dependence of the resistance and specific heat of the compound. No phase transition appears to accompany the charge correlations seen in DS experiments. Moreover, the resistance shows no abrupt changes at any temperature that can signal a transition to a long-range order CDW. The dependence of the resistivity with temperature is roughly linear with a positive slope and flattening at very low temperatures. There is a slight change in the slope of the specific heat at about 60 K, but no trace of a real phase transition. Besides, we find no signal of superconductivity in any of our measurements down to 2K [Fig. \ref{Fig3}(e)], in contrast with the 1$\mathrm{T}$-NbSeTe polymorph \cite{yan2019nbsete} and the parent compound 2$\mathrm{H}$-NbSe$_2$ \cite{Borisenko2009}. This quenching of the superconducting order in 6$\mathrm{R}$-NbSeTe concerning 2$\mathrm{H}$-NbSe$_2$ can be understood from a self-doping mechanism caused by the substitution of Se by the less electronegative Te \cite{phillips2023self}. This produces a shift in the chemical potential of the DOS of Nb atom in the $\mathrm{T}$ environment [as shown in Fig. \ref{Fig2}(i)], thus leading to a lower DOS at the Fermi level and, hence, obliterating superconductivity. Other factors like ligand disorder could also be related to the quenching, so further studies are required to elucidate this point.

\begin{figure}
\begin{center}
\includegraphics[width=\columnwidth,draft=false]{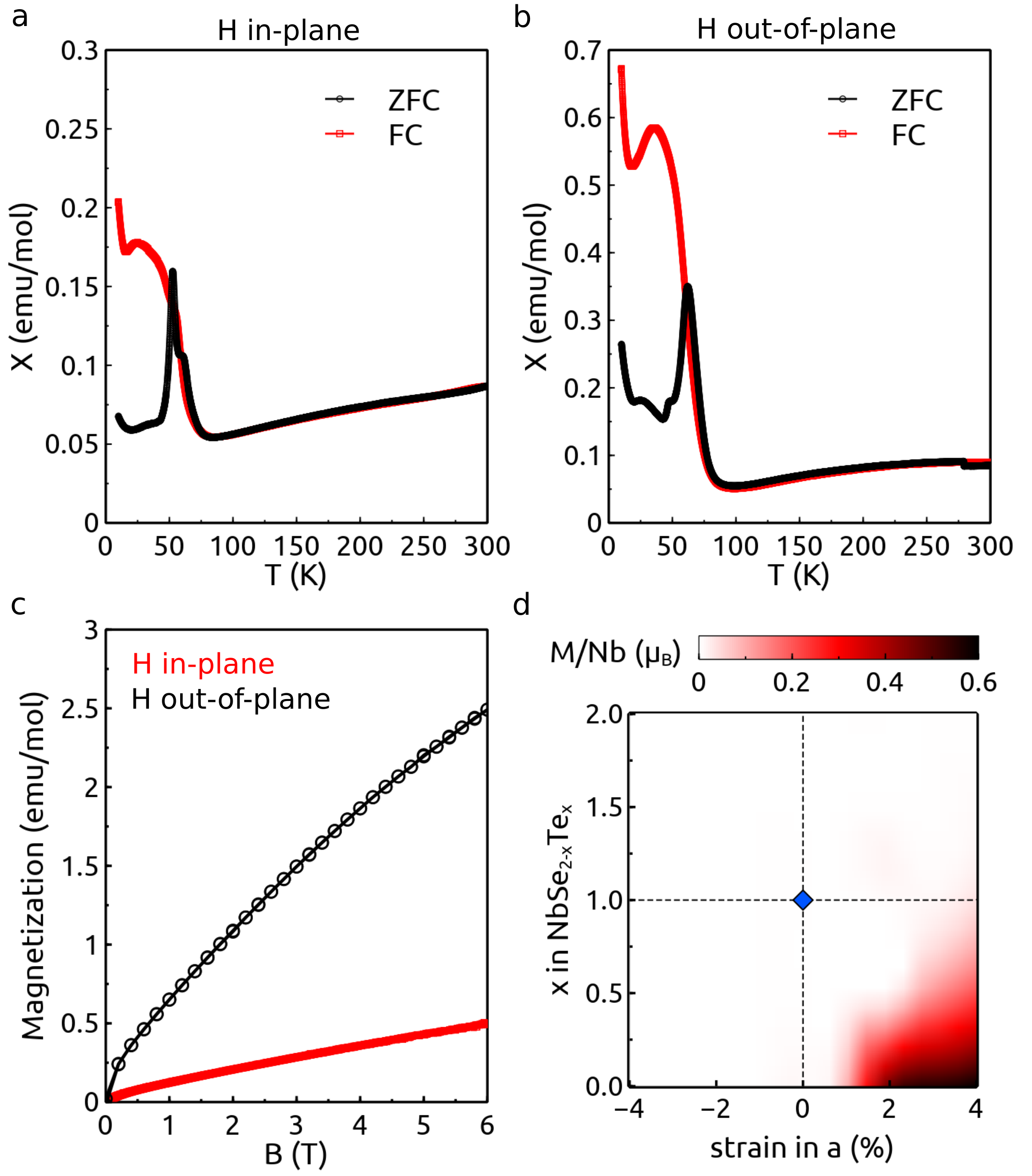}
\caption{(a) Temperature dependence of magnetic susceptibility for NbTeSe measured in the zero-field-cooled and an field-cooled condition for an applied magnetic field H= 0.5 T with H$\parallel$\textit{ab}-plane and (b) out-of-plane, H$\parallel$\textit{c}-axis. (c) Magnetization as a function of magnetic field (M \textit{vs} H) at 10 K temperature for a magnetic field applied parallel and perpendicular to the \textit{ab}-plane of the single crystal. (d) Calculated magnetic moment color map as a function of Te concentration $x$ in the Nb$_{2-x}$Te$_x$ system and strain in the in-plane lattice parameter \textit{a}. Non-zero magnetization in the supercells develops at tensile strains and low Te concentrations.
}
\label{Fig4}
\end{center}
\end{figure}

We now turn our attention to the existence of magnetism in 6$\mathrm{R}$-NbSeTe. Remarkably, the weak charge modulations are accompanied by the emergence of short-range magnetic order below $\sim$ 60 K. Figures \ref{Fig4}(a) and \ref{Fig4}(b) show the zero field cooled-field cooled (ZFC-FC) magnetic susceptibility curves of NbSeTe collected under a 0.5 T magnetic field [in-plane (a) and out-of-plane (b)]. The susceptibility initially decreases linearly from 300 K, typical of low dimensional systems, before steeply raising below 60 K. The large splitting of the ZFC-FC indicates a strongly inhomogeneous magnetic ground state \cite{Rivadulla2004} with an out-of-plane magnetic anisotropy. The inhomogeneous magnetic state is further confirmed by the absence of a $\lambda$-like anomaly in heat capacity data, Fig. \ref{Fig3}(d), and by the absence of magnetoresistance at low temperature, at about the transition temperature observed in the ZFC magnetic susceptibility. Also, the very small value of the magnetization (on the order of 10$^{-3}$ $\mu_B$/Nb) points to a glassy origin. The application of an out-of-plane magnetic field of 9T has a negligible effect on the normal state specific heat, in agreement with the absence of long-range magnetic order and the absence of linearity and saturation of the \textit{C/T} versus T$^2$ plot.   

The picture that builds up as the ground state in NbSeTe is of an entangled short-ranged charge/spin correlated ground state. To theoretically understand its microscopic origin, we have carried out DFT calculations in the 6$\mathrm{R}$ phase of NbSe$_{2-x}$Te$_{x}$ at different strain values. Figure \ref{Fig4}(d) shows the phase diagram describing the evolution of the magnetization as a function of the ligand substitution and strain. From our \textit{ab initio} calculations, we have analyzed various Te concentrations in NbSe$_{2-x}$Te$_x$ supercells at various degrees of strain. We observe that the system is typically non-magnetic, but in the tensile strain limit of the in-plane lattice parameter \textit{a} stretched by 2-4\% and low Te concentrations (x $<$ 1.0), a non-zero magnetization shows up. Given the degree of Se/Te disorder and the observed magnetic glassy behavior with a small value of the total magnetization in our samples, we interpret our experimental data on the basis of Se-rich clusters, where the local strain will produce the short-range magnetic order that gives rise to the small magnetization and ZFC-FC splitting below a certain blocking temperature.

From the CDW q-vector we have found in the DS experiments, we have carried out additional calculations in a 4 $\times$ 4 $\times$ 1 supercell to elucidate if the local-charge modulations could give rise to the emergence of magnetism. Relaxing this structure allows us to visualize in real space the striped nature (with an off-plane modulation) of this CDW (which is a lower-energy state compared to the undistorted structure). The calculations including spin polarization lead to a non-magnetic solution. Thus, the CDW itself does not bring about the observed magnetic signal. The reduction of the charge correlation length compared with the parent compounds 2$\mathrm{H}$-NbSe$_2$ and 1$\mathrm{T}$-NbTe$_2$ and the emergence of short-range magnetism show parallels to the real space spatial distribution of charge/spin and superconducting domains in superconducting cuprates \cite{Alloul2009,Suchaneck2010,Blanco2013,Campi2015}.

The question of whether the decoupled 1T-1H heterobilayers and the disordered local moment picture have anything to do with Kondo physics should be discussed, in particular given the resistivity data flattening at low temperatures. From the measurements of the specific heat, we extract a value for the Sommerfeld coefficient of about 42 mJ/mol$\cdot$K$^2$ while the value coming from our DFT calculations (at the uncorrelated GGA level) is 23 mJ/mol$\cdot$K$^2$. This kind of enhancement is typical of materials with d electrons close to the Fermi level and cannot be ascribed to Kondo physics. In our view, the magnetism observed in this system can be better explained by our calculations showing that magnetic moments may arise due to local ligand inhomogeneities and local strains in the system. This point of view is also supported by the absence of flat bands close to the Fermi level observed in the ARPES measurements (Fig. \ref{Fig2}) in contrast to the 1$\mathrm{T}$-1$\mathrm{H}$ bilayers of TaS$_2$\cite{Vano2021,crippa2023heterogeneous,Wan2023}.

\section{Conclusions}
In summary, we have presented a new type of naturally occurring NbSe$_2$-NbTe$_2$ TMD heterostructure crystallizing in the rhombohedral 6$\mathrm{R}$ phase that features a spatial distribution of short-range charge and spin orders. The combination of experimental data and \textit{ab initio} calculations have demonstrated the \textit{c}-axis stacking of $\mathrm{H}$ and $\mathrm{T}$ layers and how chemical disorder drives the presence of magnetism and electronic modulations with propagation vectors differing from the parent compounds. Our work also demonstrates the great versatility of the TMD layered structure to engineer new correlated phases of matter and opens the possibility of studying exotic ground states by means of bulk techniques (diffraction, magnetization,...), not possible for artificially stacked 2D monolayers or MBE grown TMDs.

\section*{Acknowledgements}
We thank Fernando de Juan, Eduardo da Silva Neto and Yu He for fruitful discussions and critical reading of the manuscript. This work was supported by the MINECO of Spain, projects PID2021-122609NB-C21 and PID2021-122609NB-C22. S.B-C and A.K thank the MCIN and the European Union Next Generation EU/PRTR-C17.I1, as well as by IKUR Strategy under the collaboration agreement between Ikerbasque Foundation and DIPC on behalf of the Department of Education of the Basque Government. A. O. F. thanks the financial support from the Academy of Finland Project No. 349696. J. P. thanks MECD for the financial support received through the 'Ayudas para contratos predoctorales para la formación de doctores' grant PRE2019-087338. 
We acknowledge the computational resources provided by the Galician Supercomputing Center (CESGA) and the Aalto Science-IT project. We acknowledge DESY (Hamburg, Germany), a member of the Helmholtz Association HGF, for the provision of experimental facilities. Parts of this research were carried out at PETRA III using beamline P04. Beamtime was allocated for proposal I/II-20191409. The photoemission spectroscopy instrument at beamline P04 was funded by the German Federal Ministry of Education and Research (BMBF) under the framework program ErUM (projects 05KS7FK2, 05K10FK1, 05K12FK1, 05K13FK1, 05K19FK4 with Kiel University; 05KS7WW1, 05K10WW2, and 05FK19WW2 with the University of Würzburg). Sample growth was supported by the German Research Foundation (DFG), project 434434223-SFB 1461. This work has been partly performed in the framework of the nanoscience foundry and fine analysis (NFFA-MUR Italy Progetti Internazionali) facility.

\section{Methods}

Single crystals of NbSeTe were grown by the standard chemical vapor transport method: A near-stoichiometric mixture of high-purity Nb, Se and Te was placed in a quartz ampoule together with iodine (5 mg/cm$^3$)  as transport agent; the ampoule was sealed and heated in a four-zone furnace.

Single crystal diffraction and diffuse scattering were performed at the ID28 beamline at the European Synchrotron Radiation Facility (ESRF) with \textit{E$_i$}=17.8 keV and a Dectris PILATUS3 1M X area detector. The CrysAlis software package was used for the orientation matrix refinement. Reciprocal space maps were reconstructed with ID28 software ProjectN and subsequently plotted in Albula.

Soft x-ray ARPES measurements were carried out at the P04 beamline of PETRA III at DESY using the ASPHERE photoelectron spectroscopy endstation and vacuum ultra-violet ARPES at the APE-LE beamline of ELETTRA. The soft X-ray photon energy
and total energy resolution used were 432 eV and 80 meV,
respectively. The vacuum ultra-violet photon energy and total energy resolution used were 55 eV and 30 meV, respectively. The angular resolutions were better than 0.1 degree. 

High resolution STEM imaging has been performed on TitanG2 60-300 electron microscope (FEI), equipped with x-FEG electron source, monochromator and HAADF detector. For STEM imaging the microscope was operated at 300kV at spotzise 9 and monochromator at -100V gun lens excitation below focus. Overview images were acquired in one 1Kx1K scan with 30us dwell time. Images for structure analysis were acquired as a rapid sequence of ~100 1Kx1K frames with a dwell time of 500ns, which were subsequently aligned and averaged. Projection of the unit cell has been identified and averaged over the image resulting in 126 individual averaged cells in Figure \ref{Fig1}(a).
EDX analysis was made on a probe-corrected Spectra 30-300 STEM (ThermoFisher) operated at 300 kV. Atomic resolution EDX maps were acquired with a probe current of 150 pA using Velox, and rapid rastered scanning. The Energy-Dispersive X-Ray (EDX) signal was acquired on a Dual-X system with a total acquisition angle of 1.76 Sr comprising two detectors either side of the sample.

\textit{ab initio} electronic structure calculations based on Density Functional Theory (DFT) \cite{dft} were performed using an all electron full-potential code (WIEN2k \cite{Blaha2020wien2k}). The generalized gradient approximation (GGA)\cite{perdew1996generalized} was used as the exchange-correlation term for all of our calculations. 
The harmonic phonon spectrum of the charge-density wave (CDW) phase of NbSeTe was computed using the Phonopy code \cite{phonopy}. Taking the 6R structure as the unit cell, we have performed calculations of a $4\times4$ supercell ($6\times6\times2$ \textit{k}-mesh). We have used the VASP \cite{kresse1993ab,kresse1996efficiency,kresse1996efficient} code for this task and for all the structural relaxations with the different geometries analyzed (see Suppl. Information for further details).

\section{Supplementary Material for `Self-stacked 1$\mathrm{T}$-1$\mathrm{H}$ layers in 6$\mathrm{R}$-NbSeTe and the emergence of charge and magnetic correlations due to ligand disorder'}

\subsection{Ordered ligands in the 6$\mathrm{R}$-N$\mathrm{b}$S$\mathrm{e}$T$\mathrm{e}$ unit cell}

In Supplementary Figure \ref{Fig.S1_ligand_struct}, we plot the 4 different crystalline structures used to calculate the electronic band structure and phonon dispersion of the 6$\mathrm{R}$ phases of NbSeTe. Supplementary Figure \ref{Fig.S1_ligand_struct} (a) displays a c-axis stacking of 1$\mathrm{H}$-NbSe$_2$ and 1$\mathrm{T}$-NbTe$_2$, while in Supplementary Figure \ref{Fig.S1_ligand_struct}(b), the site symmetry of Nb is swapped, namely, 1$\mathrm{T}$-NbSe$_2$ and 1$\mathrm{H}$-NbTe$_2$. Panels (c) and (d) corresponds to symmetric and anti-symmetric order of the chalcogen atoms (Se, Te) across the van der Waals gap.   

\begin{figure}
\begin{center}
\includegraphics[width=\columnwidth,draft=false]{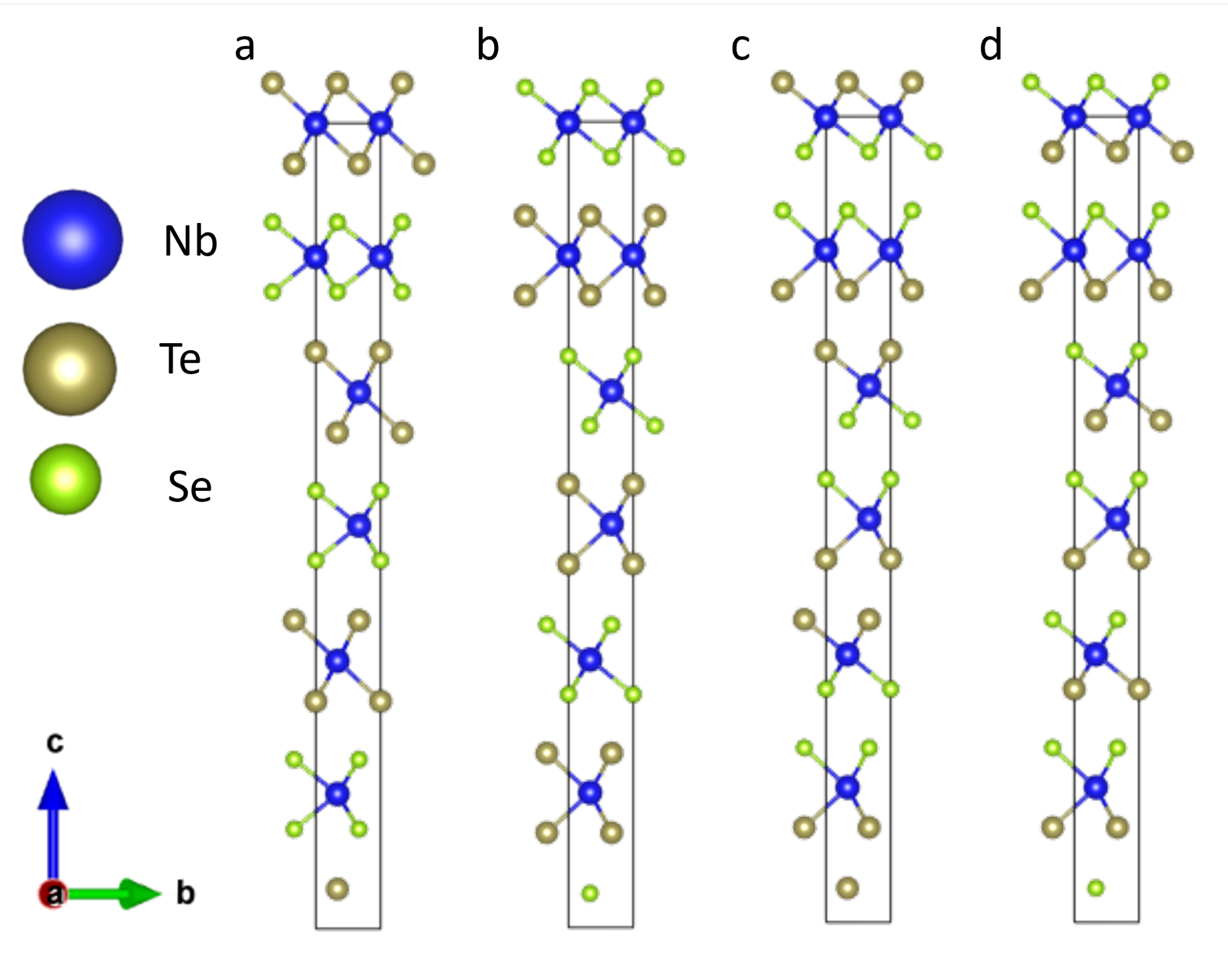}
\caption{Comparison of the different igand orderings in the 6R NbSeTe structure analyzed. Nb atoms are shown in blue, Te atoms in gold, and Se atoms in green. (a) Ordered structure with the ligands distributed as 1$\mathrm{T}$-NbTe$_2$-1$\mathrm{H}$-NbSe$_2$ and (b) 1$\mathrm{H}$-NbTe$_2$-1$\mathrm{T}$-NbSe$_2$. (c) Disordered Janus-like layers with the ligands distributed in such a way that the Se(Te) ligand atoms face Se(Te) atoms and (d) the Se(Te) ligand atoms face Te(Se) atoms.}
\label{Fig.S1_ligand_struct}
\end{center}
\end{figure} 

\subsection{Electron diffraction}

\begin{figure}
\begin{center}
\includegraphics[width=\columnwidth,draft=false]{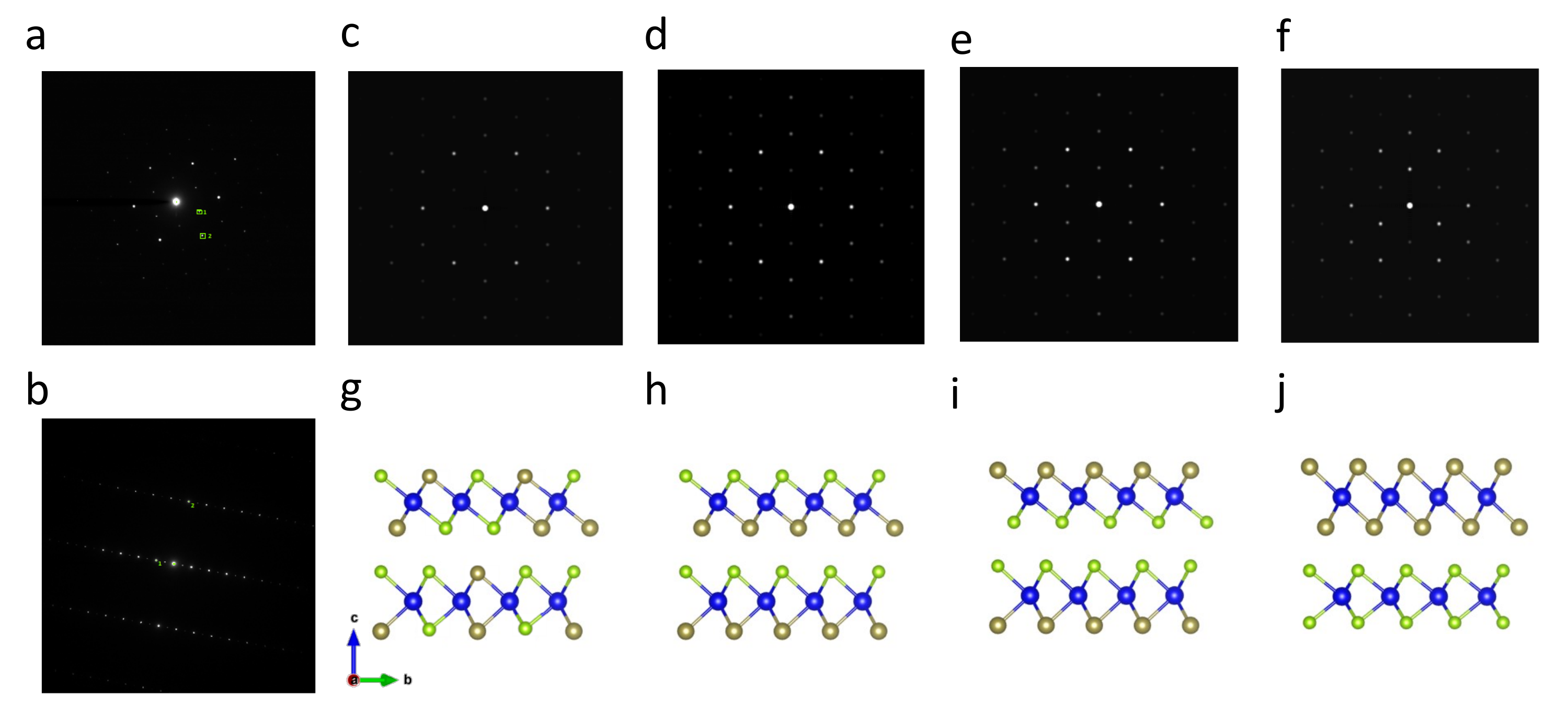}
\caption{(a) Experimental electron diffraction patterns in NbSeTe in the 001 and (b) 120 direction. Simulated electron diffraction results for the stochastic distribution (c),(g), for the disordered Janus-like layers with the ligands distributed in such a way that the Se(Te) ligand atoms face Te(Se) atoms (d),(h), for the configuration where the Se(Te) ligand atoms face Se(Te) atoms (e),(i) and for the ordered structure with the ligands distributed as 1$\mathrm{T}$-NbTe$_2$-1$\mathrm{H}$-NbSe$_2$ (f),(j).}
\label{Fig.S2_el_diff}
\end{center}
\end{figure}

Supplementary Figure \ref{Fig.S2_el_diff} displays the experimental electron diffraction, in-plane (a) and out-of-plane (b), and their in-plane simulation patterns. Supplementary Figure \ref{Fig.S2_el_diff} (c) corresponds to a stochastic distribution of Se and Te in the TMD layer with 6-fold symmetry, which nicely reproduces the experimental pattern. In Supplementary Figures \ref{Fig.S2_el_diff} (d) and (e), we plot the calculated diffraction of symmetric and anti-symmetric stacking of the TMD layers, corresponding to the stacking sequence shown in (h) and (i). All the c-orderings show a 6-fold symmetry. Nevertheless, the perfect sequence of 1H-NbSe$_2$ and 1T-NbTe$_2$ layers, corresponding to Supplementary Figure \ref{Fig.S1_ligand_struct} (a), breaks the 6-fold symmetry, hence discarding a perfect 1H-NbSe$_2$ and 1T-NbTe$_2$ ordering and favoring disorder scenarios.

\subsection{ARPES}

\begin{figure}
\begin{center}
\includegraphics[width=\columnwidth,draft=false]{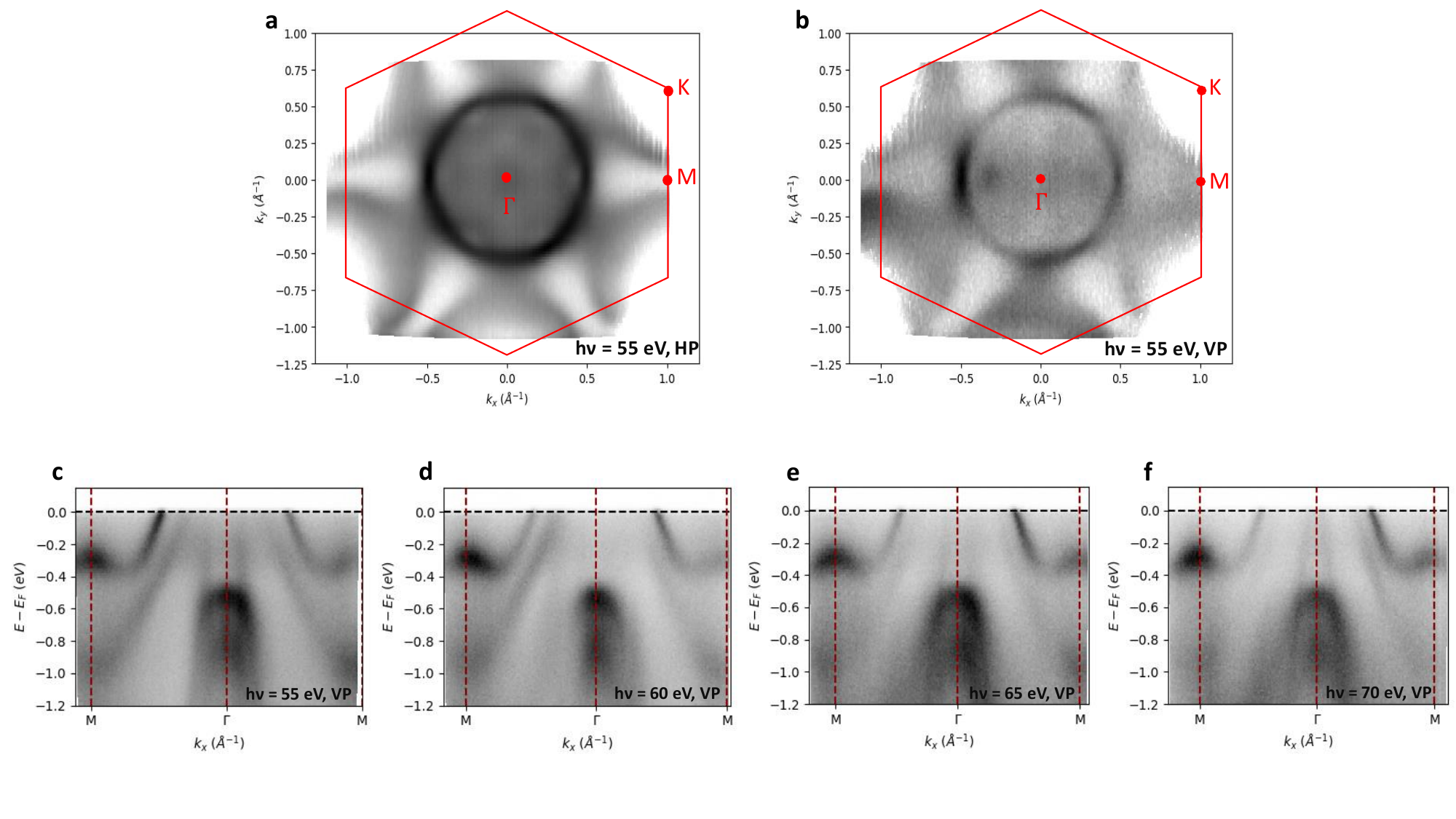}
\caption{(a) and (b) Linear horizontal polarization(HP) and vertical polarization(VP) dependent Fermi surface mapping of NbSeTe obtained with h$\nu$=55 eV and T=20 K. (c)-(f) Photon energy dependent valence band spectra of NbSeTe along the M-$\Gamma$-M symmetry direction measured with vertical polarization.
}
\label{Fig.S5_ARPES}
\end{center}
\end{figure}

Supplementary Figures \ref{Fig.S5_ARPES}(a) and \ref{Fig.S5_ARPES}(b) show the linear polarization-dependent experimental Fermi surface mapping of NbSeTe measured with h$\nu$= 55 eV and T=20 K. The red hexagon represents the first Brillouin zone of NbSeTe. Supplementary Figures \ref{Fig.S5_ARPES}(c)-(f) display photon energy
dependent valence band electronic structure of NbSeTe obtained along the M-$\Gamma$-M high-symmetry directions with vertically polarized (VP) light. Apart from intensity variation, no significant electronic structure evolution with photon energy was observed which indicates a two-dimensional character of the system.

\subsection{Diffuse scattering}

\begin{figure}
\begin{center}
\includegraphics[width=\columnwidth,draft=false]{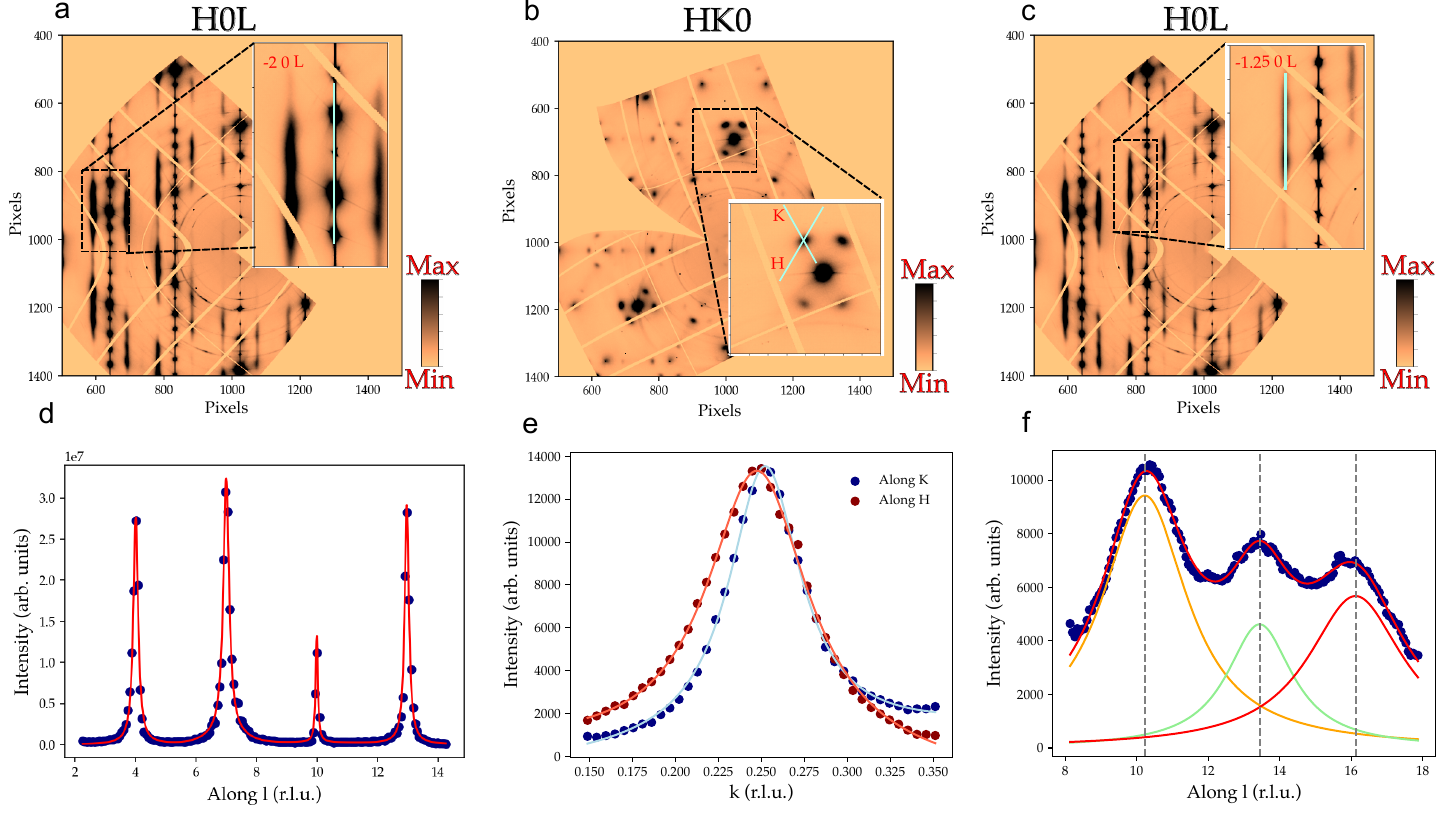}
\caption{Diffuse scattering analysis. (a) Diffuse scattering map of the  HK0 plane at $T=90\,\mathrm{K}$. (b-c) Diffuse scattering map of the plane H0L at $T=90\,\mathrm{K}$. (d) Line profile of the cuts along H and K direction along the lines in the inset of panel (a). (e) Line profile of the diffuse signal along -1.25 0 L in the inset of panel (b). The dots where fit to a sum of three Lorentzian functions. (f) Line profile along -2 0 L in the inset of the panel (c). The line shape where fit to a sum of four pseudoVoigt functions.}
\label{Fig.S3_diffuse}
\end{center}
\end{figure}

Supplementary Figure \ref{Fig.S3_diffuse} shows additional diffuse scattering data to complement that presented in the main text. In Supplementary Figure \ref{Fig.S3_diffuse}, we show the \textit{h}, \textit{k} and \textit{l} cuts of the diffuse scattering maps. In supplementary figure \ref{Fig.S3_diffuse} (a) and (b), we plot the  H 0 L map along the (0 0 1) direction and the \textit{l}-cut (b), showing the rod-like diffuse intensity along the c-direction, characteristic of a coherent 2D growth, as shown in Fig. 1 of the main text.  The H K 0 map is displayed in the Supplementary Figure \ref{Fig.S3_diffuse} (c), focusing on the area around the {1 1 0} Bragg reflection. In-plane cuts reveal a small anisotropy of the CDW peak, (see Supplementary Figure \ref{Fig.S3_diffuse} (d)). The \textit{l}-cut of the DS, with a characteristic rod-like shape, is plot in Supplementary Figure \ref{Fig.S3_diffuse}, showing a small modulation along the c-axis. 

\subsection{Structural analysis}

\setlength{\tabcolsep}{0.8em}
\begin{table}[h!]
    \centering
    \caption{Energy differences of the structures depicted in Supplementary Figure \ref{Fig.S1_ligand_struct} with respect to the ground state energy, $E_a$..}
    \label{tab:Energies}
    \begin{tabular}{|c|cccc|} \hline
         Structure & a & b & c & d   \\ \hline
         $\Delta E/ Nb\ (meV)$  & 0.0 & 93.33 & 47.67 & 28.83 \\ \hline
    \end{tabular}
\end{table}

The ab initio calculations we have performed were based on Density Functional Theory (DFT) using the Vienna Ab Initio Package (VASP) code \cite{kresse1993ab,kresse1996efficiency,kresse1996efficient}. The computed total energies used to perform energetic comparisons between the different structures analyzed were obtained setting the cutoff energy for the plane-wave-basis set (ENCUT) to 350 eV. The cut-off energy of the plane wave representation of the augmentation charges (ENAUG) was set to 500 eV. The lattice parameters and the internal positions of the atoms were fully relaxed to obtain these total energies of each structure. The converged mesh for the task was a 26$\times$26$\times$1 \textit{k}-mesh, using the Monkhorst-Pack scheme\cite{monkhorst_pack}. The harmonic phonon spectrum was computed using the real-space supercell approach\cite{phonopy, phonopy-phono3py-JPSJ}, for a 4$\times$4$\times$1 supercell and a converged 26$\times$26$\times$1 \textit{k}-mesh.

We have studied 4 different orders of the Te and Se ligands in the 6R structure. We call ordered phases to those where the $\mathrm{H}$ and $\mathrm{T}$ layers of the structure are composed of the same ligand atoms. This leads us to compare two ordered structures, one which is formed by alternating 1$\mathrm{T}$-NbTe$_2$-1$\mathrm{H}$-NbSe$_2$, as seen in Fig. \ref{Fig.S1_ligand_struct}(a), and another which is formed by alternating 1$\mathrm{H}$-NbTe$_2$-1$\mathrm{T}$- NbSe$_2$ layers, like in Fig. \ref{Fig.S1_ligand_struct}(b). We call disordered phases to those where the $\mathrm{H}$ and $\mathrm{T}$ layers of the 6R structure include both Se and Te as ligand atoms. These ligand atoms have a Janus-like order within each layer. Therefore, we have studied two disordered structures, one where the Se(Te) ligand atoms face other Se(Te) atoms of a different layer through the van der Waals (vdW) gap, as depicted in Fig. \ref{Fig.S1_ligand_struct}(c), and another where the Se(Te) ligand atoms face Te(Se) atoms of a different layer through the vdW gap, just like Fig. \ref{Fig.S1_ligand_struct}(d). For the four structures considered, we fully relaxed them (both internal coordinates and lattice parameters). The energetic comparison of these four structures can be seen in Table \ref{tab:Energies}, where the structure shown in Fig. \ref{Fig.S1_ligand_struct}(a) is the lowest energy structure and therefore is our ground state structure.  


\subsection{Charge density waves}

\begin{figure*}
\begin{center}
\includegraphics[width=\textwidth,draft=false]{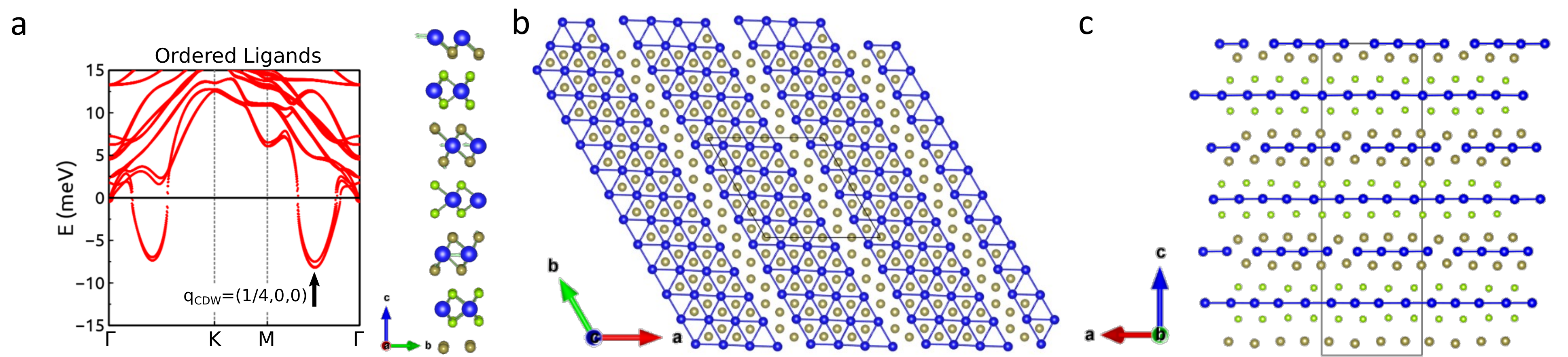}
\caption{(a) Harmonic phonon spectra for the structure in Fig. \ref{Fig.S1_ligand_struct}(a), having NbSe$_2$ (NbTe$_2$) with local 1$\mathrm{H}$ (1$\mathrm{T}$) coordination (ordered ligands), and instantaneous vibration vectors of the indicated q=(1/4, 0, 0). Scheme of the 4$\times$4$\times$1 superstructure where only the closer Nb-Nb bonds are depicted, showing the gap where the bond lengths become larger. (b) Top view of the structure in one of the layers (in this case it is irrelevant whether this is a $\mathrm{T}$- or an $\mathrm{H}$- layer. (c): Side view of the structure where the stripes run along the direction perpendicular to the paper and the off-plane phase of the stripes can be noticed.
}
\label{Fig.S4_CDW}
\end{center}
\end{figure*}

Based on the instabilities found in the harmonic phonon spectra presented in the main text, we have carried out ab initio calculations in a 4 $\times$ 4 $\times$ 1 supercell. The initial structure for the relaxation was obtained using the imaginary phonon modes in the ground state structure close to the 1/4 0 0 k-point. From Fig. \ref{Fig.S4_CDW}(a) the vibrations of this imaginary phonon mode can be seen to be mainly due to the Nb atoms in the 1-$\mathrm{T}$ coordination (surrounded by Te atoms). We created the 4$\times$4$\times$1 crystal structure using Phonopy, by displacing the atoms along normal modes at this q-point in the desired supercell. This distorted structure served as a starting point which we then relaxed. The relaxations lead to the CDW structure seen in Fig. \ref{Fig.S4_CDW}(b-c), which is lower in energy than the undistorted one 0.8 eV/Nb. We can see in Fig. \ref{Fig.S4_CDW} that this q-vector, consistent with the diffuse scattering results (see Fig. \ref{Fig.S3_diffuse}), produces a charge density wave reconstruction of the unit cell of NbSeTe that is modulated in the hexagonal plane in the form of stripes. Moreover, these stripes are out of phase in the out-of-plane direction as shown in Fig. \ref{Fig.S4_CDW}c.

\subsection{Electronic structure}

\begin{figure}
    \centering
    \includegraphics[width =\columnwidth]{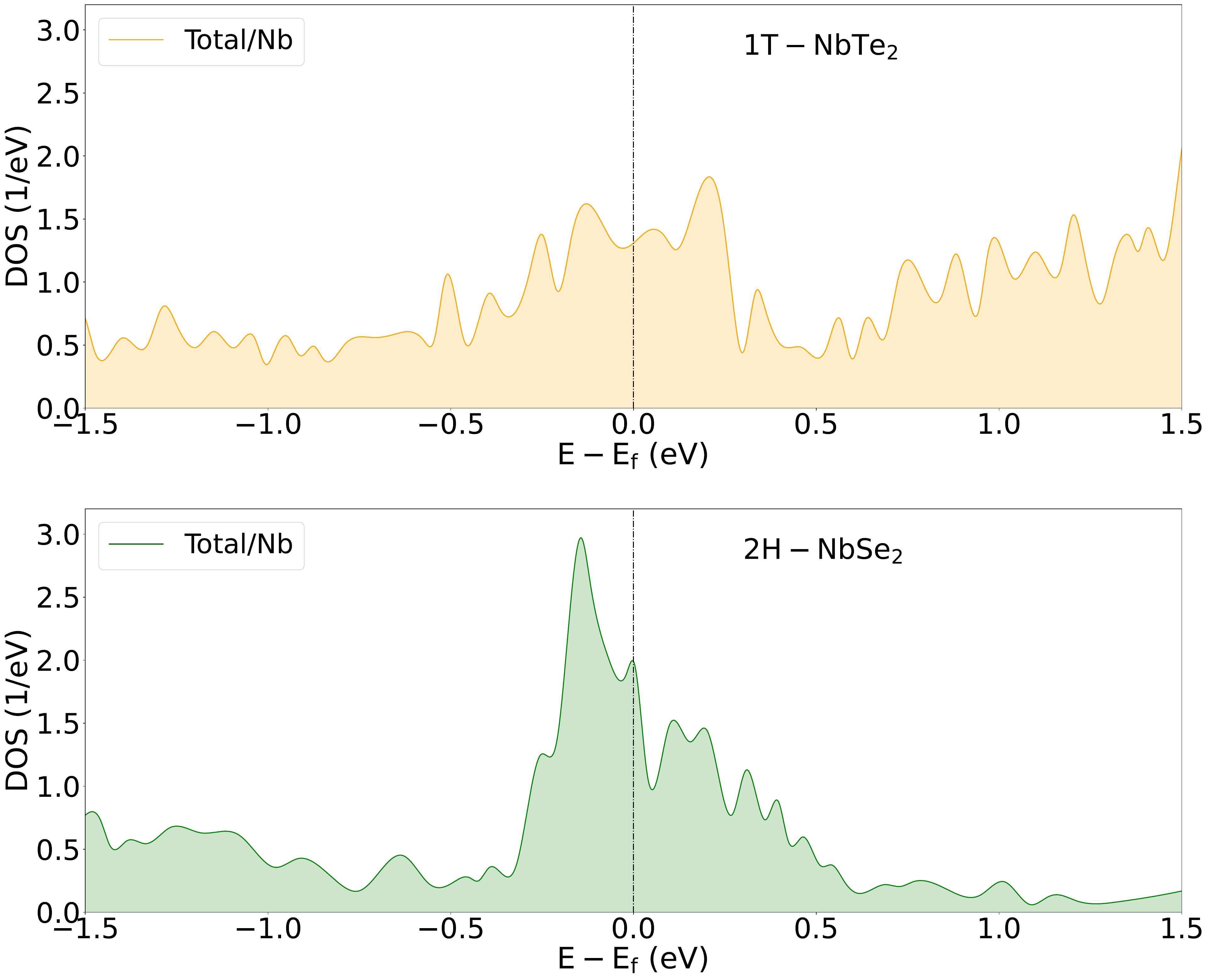}
    \caption{Density of states (DOS) of $\mathrm{1T-NbTe_2}$ (up) and $\mathrm{2H-NbSe_2}$ bulk structures with lattice parameters a and b equals to the experimental ones of NbSeTe structure and lattice parameter c being optimized for each structure. }
    \label{Fig.S6_DOS}
\end{figure}

In the main text, when discussing the density of states of the 6R supercell, we drew an approximate location for the NbSe$_2$ bulk reference chemical potential. In Supplementary Fig. \ref{Fig.S6_DOS} we present density of states calculated for the bulk 2H (1T) structure of NbSe$_2$ (NbTe$_2$) using the experimental in-plane lattice parameter of 6R-NbSeTe to draw a more direct comparison with the data of the heterostructure. We can see that the peak in DOS just below the Fermi energy characteristic of NbSe$_2$ is displaced towards the Fermi level in the bulk as compared with the 6R solution. NbSe$_2$ nominally accepts electrons when 6R-NbSeTe is formed.


%

\end{document}